\documentclass[english]{article}
\usepackage[T1]{fontenc}
\usepackage[latin9]{inputenc}
\usepackage{geometry}
\usepackage{color}
\usepackage{amsmath}
\usepackage{amsthm}
\usepackage{amssymb}
\usepackage{graphicx}
\usepackage[authoryear]{natbib}
\PassOptionsToPackage{normalem}{ulem}
\usepackage{ulem}

\makeatletter
\theoremstyle{remark}
\newtheorem{rem}{\protect\remarkname}
\theoremstyle{plain}
\newtheorem{thm}{\protect\theoremname}

\usepackage{hyperref}

\usepackage{bm} 
\newtheorem{condition}{Condition}

\makeatother

\usepackage{babel}
\providecommand{\remarkname}{Remark}
\providecommand{\theoremname}{Theorem}

\begin{document}
\title{Online False Discovery Rate Control for LORD \& SAFFRON Under Positive,
Local Dependence}
\author{Aaron Fisher\thanks{Foundation Medicine Inc.; 150 Second St, Cambridge, MA 02141}}
\date{2022-02-10}
\maketitle
\begin{abstract}
Online testing procedures assume that hypotheses are observed in
sequence, and allow the significance thresholds for upcoming tests
to depend on the test statistics observed so far. Some of the most
popular online methods include alpha investing, LORD++ (hereafter,
LORD), and SAFFRON.  These three methods have been shown to provide
online control of the ``modified'' false discovery rate (mFDR) under
a condition known as conditional superuniformity. However, to our
knowledge, LORD \& SAFFRON have only been shown to control the traditional
false discovery rate (FDR) under an independence condition on the
test statistics. Our work bolsters these results by showing that SAFFRON
and LORD additionally ensure online control of the FDR under a ``local''
form of nonnegative dependence. Further, FDR control is maintained
under certain types of adaptive stopping rules, such as stopping after
a certain number of rejections have been observed. Because alpha investing
can be recovered as a special case of the SAFFRON framework, our results
immediately apply to alpha investing as well.  In the process of
deriving these results, we also formally characterize how the conditional
superuniformity assumption implicitly limits the allowed p-value dependencies.
This implicit limitation is important not only to our proposed FDR
result, but also to many existing mFDR results.
\end{abstract}

\section{INTRODUCTION}

In traditional, prespecified hypothesis testing, all hypotheses are
known before the experiment begins. A typical goal is to limit the
probability of any false discoveries (the familywise error rate, or
FWER; see, for example \citealp{Efron2016-bj}) to be no less than
a constant denoted by $\alpha$ (e.g., $\alpha=0.05$). One simple
method for FWER control is to ``budget'' or ``spend'' a predefined
fraction of the allowed FWER on each planned test, so that the sum
of thresholds used across all tests is equal to $\alpha$. 

In contrast, \emph{online} testing procedures assume that researchers
observe a (possibly infinite) \emph{sequence} of hypotheses and associated
test statistics. Because the number of hypotheses is unknown in advance,
budgeting the alpha level across all hypotheses is not always feasible.
Instead, in a landmark paper, \citet{Foster2008-ta} propose a method
known as \emph{alpha-investing }in which rejection thresholds for
future tests are determined as a function of the statistics observed
so far. Under this framework, each rejected hypothesis replenishes
the available alpha budget, or ``alpha wealth,'' allowing users
to continue testing indefinitely.

Alpha-investing formed the inspiration for a wave of new online methods.
\citet{Aharoni2014-lv} extend this idea to ``generalized'' alpha-investing
(GAI), and \citet{Ramdas2017-rm} propose a version of GAI with improved
power (GAI++). The latter class of methods includes a special case
known as LORD++ (significance \uline{L}evels based \uline{O}n
\uline{R}ecent \uline{D}iscovery; hereafter referred to as LORD),
which is based on a method developed by \citet{Javanmard2018-rz}.
Building on LORD, \citet{Ramdas2018-qu} develop a similar method
that incorporates adaptive estimates of the proportion of true null
hypotheses, referred to as \textquotedblleft SAFFRON\textquotedblright{}
(\uline{S}erial estimate of the \uline{A}lpha \uline{F}raction
that is \uline{F}utilely \uline{R}ationed \uline{O}n true
\uline{N}ull hypotheses). 

Due to their relatively high power, LORD and SAFFRON are often described
as being among the state-of-the-art methods available for online testing
\citep{Chen2020-lj,Zhang2020-ei}. The concepts underlying these two
methods have also influenced several related online approaches \citep{Tian2019-fl,Zrnic2020-xl,Zrnic2021-jj,Xu2020-qx,Weinstein2020-gv}.
Moreover, \citet{Ramdas2018-qu} show that SAFFRON encompasses the
original alpha investing method as a special case.

Existing theoretical studies of alpha investing, LORD, and SAFFRON
typically require a condition known as \emph{conditional superuniformity}
(CS), which states that each null p-value is statistically larger
than a uniform random variable, conditional on the information used
to specify its rejection threshold (\citealp{Zrnic2021-jj}; see also
\citeauthor{Foster2008-ta,Ramdas2017-rm,Ramdas2018-qu}). Thus, each
p-value in the sequence may be ``locally dependent'' with its neighbors,
so long as its associated rejection threshold is specified sufficiently
far in advance (see details in Section \ref{subsec:Conditional-Superuniformity},
below). For example, if when several hypotheses are tested on each
of several subpopulations, then the resulting p-value sequence can
be said to follow local dependence, as tests conducted in one population
are independent of tests conducted in another. Similarly, local dependence
holds approximately if the test statistics are associated with a temporal
process with an autoregressive correlation structure. 

Under the CS assumption, LORD, and SAFFRON have been shown to control
a ``modified'' version of the false discovery rate (mFDR) \citep{Ramdas2017-rm,Ramdas2018-qu,Zrnic2021-jj}.
To our knowledge however, existing results for online control of the
\emph{traditional} false discovery rate (FDR; \citealp{Benjamini1995-sz})
require an additional assumption of p-value independence \citep{Ramdas2017-rm,Ramdas2018-qu,Zrnic2021-jj}.

Our primary contribution is to show that LORD and SAFFRON also ensure
online control of the FDR under a positive, local dependency condition
on the p-values. In particular, they control FDR whenever (1) the
null p-values are CS given the information used to define their testing
thresholds (as in \citealp{Zrnic2021-jj}); (2) the null p-values
follow the conventional assumption of positive regression dependence
on a subset (\citealp{Benjamini2001-ps}), and (3) users choose significance
thresholds that are monotonically nonincreasing in the p-values that
have been observed so far. Because alpha investing can be written
as a special case, the same results immediately apply to alpha investing
as well.

As a secondary contributions, we show that FDR control is also possible
under weakened forms of monotonicity, which allows us to account for
certain forms of adaptive stopping times. We also illustrate convenient,
special cases of the LORD and SAFFRON algorithms that may facilitate
their communication and implementation. Our proposed variations offer
added flexibility in scenarios where the logistics of hypothesis ordering
are beyond the analyst's control.

Finally, we introduce an important caveat regarding the CS assumption,
which is relevant to many results in the literature beyond our own.
For well-powered p-value sequences, we show that CS can \emph{only}
be satisfied under local dependence. To our knowledge, this is the
first formal characterization of CS as a dependence requirement. 

The remainder of this paper is organized as follows. In Section \ref{sec:Notation},
we introduce relevant notation, summarize LORD and SAFFRON, and present
existing results for FDR control. In the process, we introduce the
CS assumption and show how it implicitly restricts p-value dependencies.
Section \ref{sec:FDR-Control-our} contains our main result, and discusses
variations on the required positive dependence and monotonicity conditions.
Section \ref{sec:SIMPLIFIED-IMPLEMENTATIONS-OF} outlines convenient,
special cases of the LORD and SAFFRON algorithms. Section \ref{sec:Simulations}
illustrates our main result in a series of simulations. All proofs
are contained in the supplementary materials.

\section{NOTATION \& PRELIMINARIES\label{sec:Notation}}

Following \citet{Foster2008-ta}, we consider the setting where analysts
observe a (possibly infinite) sequence of hypotheses $\{H_{1},H_{2},\dots\}$,
along with corresponding p-values $\{P_{1},P_{2},\dots\}$. Let $\mathcal{H}_{0}\subseteq\mathbb{N}$
be the subset of indices corresponding to the truly null hypotheses.
At each stage $t$ of testing, the researcher observes $P_{t}$ and
must decide to reject or not reject $H_{t}$ before observing the
next test statistic $P_{t+1}$.

Let $\alpha_{t}$ be the significance threshold used for testing $H_{t}$,
where $H_{t}$ is rejected whenever $P_{t}\leq\alpha_{t}$. In order
to capture different types of adaptive decision making, \citet{Foster2008-ta}
allow $\alpha_{t}$ to be a function of the preceding p-values $\{P_{i}\}_{i<t}$.
To highlight that $\alpha_{t}$ may only depend on some summary function
of $\{P_{i}\}_{i<t}$, we use $\mathcal{F}_{t-1}$ to denote the minimal
set of real-valued random variables needed to determine the threshold
$\alpha_{t}$. For example, if a decision rule defines $\alpha_{t}$
only as a function of which hypotheses have been rejected\emph{ }so
far, then $\mathcal{F}_{t-1}=\{1(P_{i}<\alpha_{i})\}_{i<t}$. In this
framing, an online testing method is essentially a set of rules for
translating previous p-values into thresholds for the future tests.
Each threshold $\alpha_{t}$ is a random variable, but must be fully
determined given $\mathcal{F}_{t-1}$. 

Let $\mathcal{R}_{t}=\{i\leq t:P_{i}\leq\alpha_{i}\}$ be indices
for the hypothesis that are rejected by the $t^{th}$ stage of testing.
At each stage $t$, the false discovery proportion (FDP) and the false
discovery rate (FDR) are defined respectively as
\[
\text{FDP}(t)=\frac{\vert\mathcal{H}_{0}\cap\mathcal{R}_{k}\vert}{1\vee\vert\mathcal{R}_{k}\vert}\hspace{0.7em}\text{and}\hspace{0.7em}\text{FDR}(t)=\mathbb{E}\left[\frac{\vert\mathcal{H}_{0}\cap\mathcal{R}_{k}\vert}{1\vee\vert\mathcal{R}_{k}\vert}\right],
\]
where $(a\vee b)$ is shorthand for the maximum over $\{a,b\}$. Similarly,
the ``modified'' FDR is defined as
\[
\text{mFDR}(t)=\frac{\mathbb{E}\left[\vert\mathcal{H}_{0}\cap\mathcal{R}_{k}\vert\right]}{\mathbb{E}\left[1\vee\vert\mathcal{R}_{k}\vert\right]}.
\]

\subsection{\label{subsec:LORD-=000026-SAFFRON}LORD \& SAFFRON Approaches}

\citet{Ramdas2017-rm} suggest choosing the thresholds $\{\alpha_{i}\}_{i=1}^{\infty}$
in a way that ensures that an empirical estimate of the false discovery
proportion never exceeds $\alpha$. The authors estimate the FDP as
\[
\widehat{\text{FDP}}_{0}(t)=\frac{\sum_{i\leq t}\alpha_{i}}{1\vee\vert\mathcal{R}_{t}\vert},
\]
and propose a specific algorithm (LORD) for defining thresholds $\{\alpha_{t}\}_{t=1}^{\infty}$
so that $\widehat{\text{FDR}}_{0}(t)\leq\alpha$ for all $t\in\mathbb{N}$.
A key attribute of LORD is that it is a \emph{monotonic }algorithm,
meaning that rejecting more p-values early on can only lead to higher
testing thresholds later on (see Theorem \ref{thm:ramdas} for a formal
definition). Thus, observing a stronger signal in early tests can
only improve power for later tests. We defer the full details of the
LORD algorithm to the supplemental materials, and present a simplified,
special case of the algorithm in Section \ref{sec:SIMPLIFIED-IMPLEMENTATIONS-OF}.

Intuitively, we may expect that constraining $\widehat{\text{FDR}}_{0}(t)\leq\alpha$
will result in a small false discovery proportion, since
\begin{align}
\widehat{\text{FDP}}_{0}(t) & \geq\frac{\sum_{\{i\leq t:i\in\mathcal{H}_{0}\}}\alpha_{i}}{1\vee\vert\mathcal{R}_{t}\vert}\label{eq:alpha-frac}\\
 & \approx\frac{\sum_{\{i\leq t:i\in\mathcal{H}_{0}\}}\mathbf{1}(P_{i}\leq\alpha_{i})}{1\vee\vert\mathcal{R}_{t}\vert}\nonumber \\
 & =\text{FDP}(t).\nonumber 
\end{align}
Thus, $\widehat{\text{FDP}}_{0}(t)$ is an approximate overestimate
of $\text{FDP}(t)$, and so we can intuit that controlling $\widehat{\text{FDP}}_{0}(t)$
will result in conservative control for $\text{FDR}(t)$. Indeed,
\citet{Ramdas2017-rm} show that, under certain conditions, LORD guarantees
both $\text{mFDR}(t)\leq\alpha$ and $\text{FDR}(t)\leq\alpha$ for
all $t\in\mathbb{N}$. Importantly, these results apply not only to
the specific LORD algorithm, but to \emph{any} \emph{monotonic} algorithm
satisfying $\widehat{\text{FDP}}_{0}(t)\leq\alpha$ for all $t$ (see
details in Theorem \ref{thm:ramdas}, below). Since $\widehat{\text{FDP}}_{0}(t)$
is directly observable, the requirement that $\widehat{\text{FDP}}_{0}(t)\leq\alpha$
is straightforward to implement.

Building on this idea, \citet{Ramdas2018-qu} suggest controlling
an alternative estimate of $\text{FDP}(t)$ that is expected to be
less conservative. Leveraging strategies from \citet{Storey2002-pn}
and \citet{Storey2004-zj}, \citeauthor{Ramdas2018-qu} propose the
estimator 
\begin{equation}
\widehat{\text{FDP}}_{\mathbf{\lambda}}(t)=\frac{\sum_{i\leq t}\alpha_{i}\frac{\mathbf{1}(P_{i}>\lambda_{i})}{1-\lambda_{i}}}{1\vee\vert\mathcal{R}_{t}\vert},\label{eq:fdpl}
\end{equation}
where $\{\lambda_{i}\}_{i=1}^{\infty}$ is a series of user-defined
constants within the interval $(0,1)$. Like $\alpha_{i}$, each $\lambda_{i}$
is required to be a deterministic function of $\mathcal{F}_{i-1}$.
The intuition of $\widehat{\text{FDP}}_{\mathbf{\lambda}}(t)$ is
that $\mathbf{1}(P_{i}>\lambda_{i})/(1-\lambda_{i})$ has expectation
lower bounded by 1 when $i\in\mathcal{H}_{0}$, but has a smaller
expectation when $i\notin\mathcal{H}_{0}$. Thus, the numerator in
Eq (\ref{eq:fdpl}) will ideally have an expectation close to $\sum_{\{i\leq t:i\in\mathcal{H}_{0}\}}\alpha_{i}$,
and so $\widehat{\text{FDP}}_{\mathbf{\lambda}}(t)$ will ideally
resemble right-hand side of Line (\ref{eq:alpha-frac}). In simulations,
\citeauthor{Ramdas2018-qu} generally found setting $\lambda=1/2$
to produce algorithms with relatively high power.

\citeauthor{Ramdas2018-qu} develop a monotonic algorithm known as
SAFFRON that assigns threshold parameters $\{\alpha_{i}\}_{i=1}^{\infty}$
and tuning parameters $\{\lambda_{i}\}_{i=1}^{\infty}$ such that
$\widehat{\text{FDP}}_{\mathbf{\lambda}}(t)$ is constrained to be
no more than $\alpha$ at all stages $t\in\mathbb{N}$. Again, we
defer the full details of this algorithm to the supplemental materials,
and illustrate a simplified, special case in Section \ref{sec:SIMPLIFIED-IMPLEMENTATIONS-OF}.
\citeauthor{Ramdas2018-qu} show that, under certain conditions, any
monotonic algorithm satisfying $\widehat{\text{FDP}}_{\mathbf{\lambda}}(t)\leq\alpha$
for all $t\in\mathbb{N}$ (including SAFFRON) controls mFDR and FDR.
We review the required conditions in the next two sections.

\subsection{Conditional Superuniformity\label{subsec:Conditional-Superuniformity}}

Next, we study the conditional superuniformity (CS) assumption, which
forms the basis for many procedures in the online testing field.\emph{
}Formally,\emph{ }we say that a p-value $P_{t}$\emph{ }satisfies
CS if $\mathbb{P}(P_{t}\leq u|\mathcal{F}_{t-1})\leq u$, i.e., $P_{t}$
is a valid p-value even conditional on the information used to define
its rejection threshold, $\mathcal{F}_{t-1}$. This assumption is
nontrivial to verify: if joint distribution of underlying test statistics
is completely unknown, then there is no clear way to produce p-values
that are CS. 

\citet{Zrnic2021-jj} propose a clever means of circumventing this
problem when\emph{ partial knowledge }of\emph{ }this joint distribution
is available. In particular, the authors consider cases where each
p-value is dependent only with a subset of its neighbors. Reflecting
these local relationships, we will refer to a p-value $P_{t}$ as
following \emph{local dependence} whenever $P_{t}$ is independent
of the information ($\mathcal{F}_{t-1}$) used to define its thresholds
$(\alpha_{t},\lambda_{t})$.f For example, suppose that test statistics
are observed in batches, and are known to be independent across batches.
Let $b_{t}$ be the batch label for the $t^{th}$ hypothesis. Even
if the within-batch dependencies are unknown, we can still proceed
by choosing parameters for upcoming tests only based on the test statistics
from previous batches. By constraining $\mathcal{F}_{t-1}$ to be
a function of $\{P_{i}:b_{i}<b_{t}\}$, we can effectively ignore
the within-batch dependencies, as $\mathbb{P}(P_{t}\leq\alpha_{t}\vert\mathcal{F}_{t-1})=\mathbb{P}(P_{t}\leq\alpha_{t})$. 

While CS is often implied by local dependence, it is not immediately
obvious whether the reverse is true. A more precise understanding
of how CS restricts dependencies does not appear to have been illustrated
in the literature. To explore this link further, we propose the following,
novel remark.
\begin{rem}
\emph{(CS Necessary Conditions)\label{rem:cs-requirements}If $P_{t}$
is continuous, then $P_{t}$ can satisfy CS only if one of the following
two conditions holds.}
\begin{enumerate}
\item \emph{($P_{t}$ follows locally dependence) $P_{t}\perp\mathcal{F}_{t-1}$.\label{enu:(-follows-locally}}
\item \emph{($P_{t}$ is underpowered) There exists an alternative p-value
$P_{t}'$ satisfying both CS and local dependence, and that is strictly
more powerful than $P_{t}$. More specifically, there exists a random
variable $P_{t}'$ satisfying $\mathbb{P}(P_{t}'\leq u|\mathcal{F}_{t-1})=u$;
$P_{t}'\perp\mathcal{F}_{t-1}$; $P_{t}'\leq P_{t}$}\textcolor{black}{\emph{;
and $\mathbb{P}(P_{t}'<P_{t})>0$.}}\emph{ \label{enu:(-is-underpowered)}}
\end{enumerate}
\end{rem}
The implication of the Remark \ref{rem:cs-requirements} is that,
in order to achieve CS, we must either plan testing thresholds for
$P_{i}$ only using variables that are uninformative of $P_{i}$ (as
in \citealp{Zrnic2021-jj}), or we must use p-values that are underpowered.
That is, for well-powered p-value sequences, local dependence is a\emph{
necessary} \emph{condition} for CS.

\subsection{Existing FDR Bounds for LORD and SAFFRON, Under Independence\label{sec:Existing-FDR-bounds}}

\citet{Zrnic2021-jj} show that LORD and SAFFRON each control mFDR
under the CS condition (see their Theorem 2; as well as \citealp{Ramdas2017-rm};
and \citealp{Ramdas2018-qu}).
\begin{thm}
\label{thm:mfdr}(mFDR under CS) Assume that all p-values satisfy
CS (i.e., $\mathbb{P}(P_{t}\leq u|\mathcal{F}_{t-1})$ for all $t\in\mathbb{N}$),
and that $\alpha_{t}$ and (when applicable) $\lambda_{t}$ are deterministic
functions of $\mathcal{F}_{t-1}$. Under these conditions, the following
two results hold.
\begin{enumerate}
\item (LORD mFDR Control) If the parameters $\{\alpha_{i}\}_{i\in\mathbb{N}}$
are selected so that $\widehat{\text{FDR}}_{0}(t)\leq\alpha$ for
all $t\in\mathbb{N}$ (e.g., LORD) then $\text{mFDR}(t)\leq\alpha$. 
\item (SAFFRON FDR Control) If the parameters $\{\alpha_{i},\lambda_{i}\}_{i\in\mathbb{N}}$
are selected so that $\widehat{\text{FDR}}_{\lambda}(t)\leq\alpha$
for all $t\in\mathbb{N}$ (e.g., SAFFRON) then $\text{mFDR}(t)\leq\alpha$.
\end{enumerate}
\end{thm}
Additionally, \citet{Ramdas2017-rm} and \citet{Ramdas2018-qu} show
that traditional FDR control is achieved under a combination of CS,
a monotonicity condition, and a p-value independence condition.
\begin{thm}
\label{thm:ramdas}(FDR under independence) In addition to the conditions
of Theorem \ref{thm:mfdr}, we make the following assumptions.
\begin{enumerate}
\item (Independence) The null p-values are independent of each other and
the non-nulls; 
\item \label{cond:monotonic-saff}(Monotonicity) For each $t\in\mathbb{N}$,
the parameters $\alpha_{t}$ and (when applicable) $\lambda_{t}$
are deterministic, coordinatewise nondecreasing functions of the set
of indicators $\mathcal{F}_{t-1}:=(\mathbf{1}(P_{1}\leq\alpha_{1}),\dots,\mathbf{1}(P_{t-1}\leq\alpha_{t-1}),\mathbf{1}(P_{1}\leq\lambda_{1}),\dots,\mathbf{1}(P_{t-1}\leq\lambda_{t-1}))$.
Note that, for LORD, this condition can be simplified by fixing each
$\lambda_{t}=0$.
\end{enumerate}
Under the above conditions, the following two results hold.
\begin{enumerate}
\item (LORD FDR Control) If the parameters $\{\alpha_{i}\}_{i\in\mathbb{N}}$
are selected so that $\widehat{\text{FDR}}_{0}(t)\leq\alpha$ for
all $t\in\mathbb{N}$ (e.g., LORD) then $\text{FDR}(t)\leq\alpha$. 
\item (SAFFRON FDR Control) If the parameters $\{\alpha_{i},\lambda_{i}\}_{i\in\mathbb{N}}$
are selected so that $\widehat{\text{FDR}}_{\lambda}(t)\leq\alpha$
for all $t\in\mathbb{N}$ (e.g., SAFFRON) then $\text{FDR}(t)\leq\alpha$.
\end{enumerate}
\end{thm}
The monotonicity condition in Theorem \ref{thm:ramdas} essentially
states that lower p-values never lead us to require stricter thresholds
in future tests -- the more hypotheses we reject, the easier it will
be to reject future hypotheses. This monotonicity condition can be
ensured by design. 

In the next section, we show that the additional assumptions made
in Theorem \ref{thm:ramdas} can be greatly relaxed. Most notably,
the independence assumption can be weakened to a positive dependence
assumption. That said, we still require conditional superuniformity.

\section{FDR CONTROL UNDER POSITIVE DEPENDENCE\label{sec:FDR-Control-our}}

In Section \ref{subsec:Main-result}, below, we introduce a monotonicity
condition and a nonnegative dependence condition. We then present
our main result, along with a sketch of the proof. In Sections \ref{subsec:Incorporating-certain-forms},
we discuss how certain forms of adaptive stopping rules can be formalized
under our assumptions.

\subsection{Main Result\label{subsec:Main-result}}

Our first required condition for online FDR control is analogous to
the monotonicity conditions described in Theorem \ref{thm:ramdas},
and can similarly be ensured by design. We will require that any decrease
to a p-value (i.e., making it ``more significant'') cannot decrease
the total number of rejections. 

\begin{condition}

\label{cond:my-mono} (Relaxed Monotonicity) For any $t\in\mathbb{N}$
and for any two p-value vectors $\mathbf{p}=(p_{1},\dots,p_{t})$
and $\mathbf{p}'=(p_{1}',\dots,p_{t}')$ that could possibly result
from the first $t$ tests, if $p_{i}\leq p_{i}'$ for all $i\leq t$,
then $\mathbf{p}$ must produce at least as many rejections as $\mathbf{p}'$.

\end{condition}

The most straightforward way to ensure Condition \ref{cond:my-mono}
is to require that each threshold $\alpha_{t}$ be a monotonic, nonincreasing
function of the preceding p-values $(P_{1},\dots,P_{t-1})$.\footnote{Note that any nonincreasing function of the p-values is \emph{nondecreasing}
in the \emph{rejection indicators} $\mathcal{L}_{t-1}$.} However, we will see in the next section that weaker versions of
monotonicity can also satisfy Condition \ref{cond:my-mono}.

In order to relax the independence requirement in Theorem \ref{thm:ramdas},
we next introduce a version of the well-known \emph{positive regression
dependence on a subset }(PRDS) condition developed by \citet{Benjamini2001-ps}.
This condition will depend on the notion of \emph{increasing sets.
}A set of $k$-dimensional vectors $D\in[0,1]^{k}$ is called increasing
if, for any vector $\mathbf{x}=(x_{1},\dots,x_{t})\in D$ and any
vector $\mathbf{y}=(y_{1},\dots,y_{t})$ satisfying $x_{i}\leq y_{i}$
for all $i$, it must also be that $\mathbf{y}\in D$. For example,
for any $r\leq t$, Condition \ref{cond:my-mono} implies that the
set of p-values that produce no more than $r$ rejections by stage
$t$ is an increasing set. 

\begin{condition}

\label{cond:my-PRDS} (Conditional PRDS) For any stage $t$, any
null index $i\leq t$ satisfying $i\in\mathcal{H}_{0}$, and increasing
set $D\subset[0,1]^{t}$, the probability $\mathbb{P}((P_{1},\dots,P_{t})\in D|P_{i}=u,\mathcal{F}_{i-1})$
is nondecreasing in $u$. 

\end{condition}

Roughly speaking Condition \ref{cond:my-PRDS} says that each null
p-value is positively associated with the other p-values. As with
local dependence, we may expect Condition \ref{cond:my-PRDS} to hold
when sequentially analyzing population subgroups, with several hypotheses
tested per group. We may also expect Condition \ref{cond:my-PRDS}
to hold when studying temporal test statistics believed to follow
an autoregressive structure.

The key take-away from Conditions \ref{cond:my-mono} \& \ref{cond:my-PRDS}
is that, together, they imply that low p-values in early stages of
the sequence tend to be associated with a higher number of rejections
by later stages of the sequence. This idea will play a central role
in our derivation of FDR control.

\begin{thm}
\label{thm:dep}(FDR under nonnegative dependence) Under Conditions
\ref{cond:my-mono} \& \ref{cond:my-PRDS}, if the p-values are conditionally
superuniform (i.e., $\mathbb{P}(P_{t}\leq u|\mathcal{F}_{t-1})\leq u)$
for all $t\in\mathbb{N}$), then the following two results hold.
\begin{enumerate}
\item (LORD FDR Control) If the parameters $\{\alpha_{i}\}_{i\in\mathbb{N}}$
are selected so that $\widehat{\text{FDR}}_{0}(t)\leq\alpha$ for
all $t\in\mathbb{N}$ (e.g., LORD) then $\text{FDR}(t)\leq\alpha$. 
\item (SAFFRON FDR Control) If the parameters $\{\alpha_{i},\lambda_{i}\}_{i\in\mathbb{N}}$
are selected so that $\widehat{\text{FDR}}_{\lambda}(t)\leq\alpha$
for all $t\in\mathbb{N}$ (e.g., SAFFRON) then $\text{FDR}(t)\leq\alpha$.
\end{enumerate}
\end{thm}
Importantly, Theorem \ref{thm:dep} still requires the CS condition.
Thus, for well-powered p-value sequences, Theorem \ref{thm:dep} effectively
requires the p-values satisfy nonnegative, local dependence (see Section
\ref{subsec:Conditional-Superuniformity}).

A sketch of the intuition for Theorem \ref{thm:dep} is as follows.
Consider the especially simple case where all parameters $\{\alpha_{i},\lambda_{i}\}_{i\in\mathbb{N}}$
are fixed a priori (i.e., $\mathcal{F}_{i-1}=\emptyset$ for all $i$).
This setting greatly simplifies the notation needed, and still sheds
light on how Theorem \ref{thm:dep} can be proved when $\alpha_{i}$
and $\lambda_{i}$ are determined adaptively (see comments below).
Under Conditions \ref{cond:my-mono} \& \ref{cond:my-PRDS}, smaller
null p-values are generally associated with \emph{larger} values for
$(1\vee\vert\mathcal{R}_{t}\vert)$. Thus, for any $i\in\mathcal{H}_{0}$
and $t\geq i$, we would expect the rejection indicator $\mathbf{1}(P_{i}\leq\alpha_{i})$
to be \emph{negatively correlated} with $1/(1\vee\vert\mathcal{R}_{t}\vert)$,
i.e.,
\begin{equation}
\mathbb{E}\left[\frac{1}{1\vee\vert\mathcal{R}_{t}\vert}\mathbf{1}(P_{i}\leq\alpha_{i})\right]\leq\mathbb{E}\left[\frac{1}{1\vee\vert\mathcal{R}_{t}\vert}\right]\mathbb{E}\left[\mathbf{1}(P_{i}\leq\alpha_{i})\right].\label{eq:neg-cor}
\end{equation}
Applying this, we have
\begin{align}
\text{FDR}(t)= & \sum_{\{i\leq t;i\in\mathcal{H}_{0}\}}\mathbb{E}\left[\frac{\mathbf{1}(P_{i}\leq\alpha_{i})}{1\vee\vert\mathcal{R}_{t}\vert}\right]\nonumber \\
\leq & \sum_{\{i\leq t;i\in\mathcal{H}_{0}\}}\mathbb{E}\left[\frac{1}{1\vee\vert\mathcal{R}_{t}\vert}\right]\mathbb{E}\left[\mathbf{1}(P_{i}\leq\alpha_{i})\right]\nonumber \\
\leq & \sum_{\{i\leq t;i\in\mathcal{H}_{0}\}}\mathbb{E}\left[\frac{1}{1\vee\vert\mathcal{R}_{t}\vert}\right]\alpha_{i}\nonumber \\
= & \mathbb{E}\left[\sum_{\{i\leq t;i\in\mathcal{H}_{0}\}}\frac{\alpha_{i}}{1\vee\vert\mathcal{R}_{t}\vert}\right],\label{eq:fdr-limit-simple}
\end{align}
where the first inequality comes from Eq (\ref{eq:neg-cor}), and
the second inequality comes from $\mathbb{P}(P_{i}\le\alpha_{i})=\mathbb{P}(P_{i}\le\alpha_{i}\vert\mathcal{F}_{i-1})\leq\alpha_{i}$.
Thus, if $\widehat{\text{FDP}}_{0}(t)=\sum_{i\leq t}\frac{\alpha_{i}}{\vert1\vee\mathcal{R}_{t}\vert}\leq\alpha$,
then monotonicity of expectations implies that $\text{FDR}(t)\leq\alpha$. 

To sketch the result for $\widehat{\text{FDP}}_{\lambda}$, we build
on Eq (\ref{eq:fdr-limit-simple}) by multiplying each summation term
by $\mathbb{E}\left[\mathbf{1}(P_{i}>\lambda_{i})\right]/(1-\lambda)\geq1$.
We obtain 
\begin{align*}
\text{FDR}(t) & \leq\sum_{\{i\leq t;i\in\mathcal{H}_{0}\}}\mathbb{E}\left[\frac{\alpha_{i}}{1\vee\vert\mathcal{R}_{t}\vert}\right]\frac{\mathbb{E}\left[\mathbf{1}(P_{i}>\lambda_{i})\right]}{1-\lambda_{i}}\\
 & \leq\sum_{\{i\leq t;i\in\mathcal{H}_{0}\}}\mathbb{E}\left[\frac{\alpha_{i}\frac{\left[\mathbf{1}(P_{i}>\lambda_{i})\right]}{1-\lambda_{i}}}{1\vee\vert\mathcal{R}_{t}\vert}\right],
\end{align*}
where the second inequality comes from the fact that indicators of
\emph{large }p-values, $\mathbf{1}(P_{i}>\lambda_{i})$, are \emph{positively}
\emph{correlated }with $\left(1\vee\vert\mathcal{R}_{t}\vert\right)^{-1}$.
From here, if $\widehat{\text{FDP}}_{\lambda}(t)\leq\alpha$, then
monotonicity of expectations again implies that $\text{FDR}(t)\leq\alpha$.
In the full details of the proof, we also iterate expectations over
$\mathcal{F}_{i-1}$ in order to account for adaptively defined parameters
$\alpha_{i}$ and $\lambda_{i}$ (see the supplemental materials).

\subsection{Incorporating Certain Forms of Adaptive Stopping Rules\label{subsec:Incorporating-certain-forms}}

While Theorem \ref{thm:dep} ensures FDR control at fixed times $t\in\mathbb{N}$,
it does not \emph{uniformly} control the FDR across all times $t$.
That is, we cannot conclude from Theorem \ref{thm:dep} that $\sup_{t\in\mathbb{N}}\text{FDR}(t)\leq\alpha$.
The practical relevance of this point is that some analysts may choose
their final test stage \emph{adaptively}, in the hopes rejecting a
large proportion of the hypotheses tested. In other words, they may
wish to stop testing early in the face of especially strong preliminary
results. We show in this section that, for certain types of adaptive
stopping rules, the conditions required for Theorem \ref{thm:dep}
still hold.

We will use $T$ to denote an adaptively determined stopping time,
and will say that FDR is controlled under adaptive stopping times
if $\mathbb{E}\left[\text{FDR}(T)\right]\leq\alpha.$ We will generally
assume that $\mathbf{1}(T>t)$ is a deterministic function of $(P_{1},\dots,P_{t-1})$,
and that $T$ is upper bounded (with probability 1) by a known constant
$t_{\text{max}}$.

At first glance, it appears straightforward to account for these kinds
of adaptive stopping times by simply setting $\alpha_{t}$ equal to
zero for every stage $t>T$, and continuing testing until stage $t_{\text{max}}$.
By controlling $\text{FDR}(t_{\text{max}})$, we effectively control
$\mathbb{E}\left[\text{FDR}(T)\right]$. Unfortunately, such a method
of assigning thresholds is \emph{not monotonic} in the observed p-values.
Seeing a sufficient number of rejections may cause us to \emph{lower}
our thresholds for all future tests by setting them to zero, and so
the monotonicity conditions in Theorem \ref{thm:ramdas} are not satisfied.

However, even though these kinds of stopping rules are not compatible
with monotonicity constraints in Theorem \ref{thm:ramdas}, there
are several types of adaptive stopping rules that still maintain Condition
\ref{cond:my-mono}. In particular, Condition \ref{cond:my-mono}
may still be satisfied if users combine a monotone threshold function
with a monotone stopping rule. To formalize these types of rules,
let $\{\beta_{i}\}_{i=1}^{\infty}$ be a sequence of functions used
for defining thresholds, where $\beta_{i}:[0,1]^{i-1}\rightarrow[0,1]$
is a mapping from first $i-1$ p-values to the threshold $\alpha_{i}=\beta_{i}(P_{1},\dots,P_{i-1})$. 

As a first example, consider the case where users assign each threshold
$\alpha_{i}$ according to a coordinatewise nonincreasing function
$\beta_{i}^{\text{noninc}},$ but can also decide to stop testing
if a certain critical threshold $\gamma^{\text{max-R}}$ of rejections
have been made, or if a certain number of tests $\gamma^{\text{max-stage}}$
have completed. This implies the following form for each function
$\beta_{i}$
\begin{align*}
\beta_{i}(P_{1},\dots,P_{i-1}) & =\beta_{i}^{\text{noninc}}(P_{1},\dots,P_{i-1})\\
 & \hspace{1em}\times\mathbf{1}\left(\vert\mathcal{R}_{i-1}\vert<\gamma^{\text{max-R}}\right)\\
 & \hspace{1em}\times\mathbf{1}\left(i\leq\gamma^{\text{max-stage}}\right).
\end{align*}
Here, the overall function $\beta_{i}$ is not monotonic -- smaller
p-values will increase future thresholds at first, but eventually
will cause them to drop to zero. Still, even though the thresholds
themselves are not monotonic in the p-value sequence, the total number
of \emph{rejections} \emph{is} monotonic in the p-value sequence.
Any decrease to a p-value can lead to a cease of testing, but cannot
decrease the total number of rejections, and so Condition \ref{cond:my-mono}
still holds.

A similar result holds if we allow the maximum number of rejections,
or the maximum number of stages, to be extended in the face of strong
signal in the preliminary tests. To formalize this, we can instead
assume that each function $\beta_{i}$ has the following structure.
\begin{align*}
\beta_{i}(P_{1},\dots,P_{i-1}) & =\beta_{i}^{\text{noninc}}(P_{1},\dots,P_{i-1})\\
 & \hspace{1em}\times\mathbf{1}\left(\vert\mathcal{R}_{i-1}\vert<\beta^{\text{max-R}}(P_{1},\dots,P_{i-1})\right)\\
 & \hspace{1em}\times\mathbf{1}\left(i\leq\beta^{\text{max-stage}}(P_{1},\dots,P_{i-1}\right),
\end{align*}
where $\beta_{i}^{\text{max-R}}:[0,1]^{i-1}\rightarrow\mathbb{N}$
and $\beta_{i}^{\text{max-stage}}:[0,1]^{i-1}\rightarrow\mathbb{N}$
are nonincreasing functions of $(P_{1},\dots,P_{i-1})$. Here, roughly
speaking, larger p-values in the early stages must produce either
stricter thresholds $($via $\beta_{i}^{\text{noninc}}$), a lower
number of maximum stages (via $\beta_{i}^{\text{max-stage}})$, or
a lower number of maximum rejections (via $\beta_{i}^{\text{max-R}})$.
Any of these changes can only reduce the number of total discoveries.
We again see that Condition \ref{cond:my-mono} holds: the threshold
functions $\beta_{i}$ are not themselves monotonic, but the number
of discoveries is still nonincreasing in the observed p-values.

\section{CONVENIENT IMPLEMENTATIONS OF LORD \& SAFFRON\label{sec:SIMPLIFIED-IMPLEMENTATIONS-OF}}

Both LORD \& SAFFRON require the user to specify a threshold $\alpha_{1}$
for the first test statistic $P_{1}$, as well as a sequence of tuning
parameters $\{\gamma_{i}\}_{i=1}^{\infty}$ summing to one. If users
place more weight on the early terms of this sequence, then LORD and
SAFFRON will allocate more alpha wealth to the early stages of testing,
as well as to tests immediately following a discovery. 

While the details of these two algorithms are fairly involved, we
remark in this section that they can be greatly simplified whenever
$\{\gamma_{i}\}_{i=1}^{\infty}$ is chosen to be a geometric series.
To our knowledge, the convenient forms of the LORD and SAFFRON algorithms
discussed below have not been illustrated before.

We also include simplified versions of the algorithms outlined by
\citet{Zrnic2021-jj}, which are designed to account for various forms
of partially unknown dependencies among the test statistics (see summary
in Section \ref{subsec:Conditional-Superuniformity}). In doing so,
we also relax one of the ``monotonicity'' assumptions made by \citeauthor{Zrnic2021-jj}.

\subsection{Base implementations\label{subsec:Base-implementations}}

We define $\pi\in[0,1]$ to be a user-specified constant representing
the proportion of the available ``alpha wealth'' to be allocated
to the current test. We show in the supplemental materials that, if
we choose tuning parameters $\alpha_{1}=\pi\alpha$ and $\gamma_{i}=\pi(1-\pi)^{i-1}$
for $i\in\mathbb{N}$, then the LORD algorithm becomes equivalent
to setting
\begin{equation}
\alpha_{t}=\left\{ \alpha\left(1\vee\vert\mathcal{R}_{t-1}\vert\right)-\sum_{i\leq t-1}\alpha_{i}\right\} \pi\label{eq:lord-simple}
\end{equation}
for all $t\in\mathbb{N}$, where $\mathcal{R}_{0}=\emptyset$. It
is readily apparent that defining $\alpha_{t}$ in this way ensures
that $\widehat{\text{FDR}}_{0}(t)\leq\alpha$ for all $t\in\mathbb{N}$. 

The SAFFRON algorithm described by \citet{Ramdas2018-qu} additionally
requires users to specify a constant $\lambda\in(0,1)$, and sets
$\lambda_{i}=\lambda$ for all $i\in\mathbb{N}.$ Under the same choice
of tuning parameters as above (see details in the supplemental materials),
SAFFRON similarly reduces to setting setting each $\alpha_{t}=\min\{\lambda_{t},\bar{\alpha}_{t}\}$,
where
\begin{align}
\bar{\alpha}_{t} & =\left\{ \alpha\left(1\vee\vert\mathcal{R}_{t-1}\vert\right)-\sum_{i\leq t-1}\frac{\bar{\alpha}_{i}\mathbf{1}(P_{i}>\lambda)}{1-\lambda}\right\} \left(1-\lambda\right)\pi.\label{eq:saff-simple}
\end{align}
By rearranging terms, we can see that this choice of $\{\lambda_{t},\alpha_{t}\}_{t=1}^{\infty}$
indeed ensures that $\widehat{\text{FDP}}_{\mathbf{\lambda}}(t)\leq\alpha$
for all $t\in\mathbb{N}$. SAFFRON's use of $\min\{\lambda,\bar{\alpha}_{t}\}$
as a threshold, rather than $\bar{\alpha}_{t}$, can be motivated
as maintaining the intuition that p-values larger than $\lambda$
are likely associated with true null hypotheses, and should therefore
not be rejected. Of course, $\widehat{\text{FDP}}_{\mathbf{\lambda}}(t)$
remains controlled if we replace $\bar{\alpha}_{i}$ with $\alpha_{i}$
in the right-hand side of Eq (\ref{eq:saff-simple}).

\subsection{Incorporating planning ahead\label{subsec:planning-ahead-implementation}}

Next, we generalize Eqs (\ref{eq:lord-simple}) \& (\ref{eq:saff-simple})
to account for the kinds of information restrictions studied by \citeauthor{Zrnic2021-jj}
(\citeyear{Zrnic2021-jj}, see Section \ref{subsec:Conditional-Superuniformity}
for a summary). Let $s_{i}<i$ denote the time by which $\alpha_{i}$
and $\lambda_{i}$ must be specified. For example, $s_{i}=0$ for
all $i\in\mathbb{N}$ recovers the ``alpha spending'' setting in
which all thresholds must be completely prespecified, and $s_{i}=i-1$
for all $i\in\mathbb{N}$ recovers the canonical online setting of
Theorem \ref{thm:ramdas}. Alternatively, $s_{i}$ may denote the
last stage before the ``batch'' that contains $H_{i}$, or, more
generally, the last stage $k$ satisfying $(P_{1},\dots,P_{k})\perp P_{i}$.
We can also use the same notation ($s_{i}$) to describe settings
where the parameters for testing $H_{i}$ must be specified several
stages before the result is announced (at stage $i$) due to logistical
delays in the scientific process. \citeauthor{Zrnic2021-jj} discuss
all of these framings in detail. 

One straightforward way to ensure that $\widehat{\text{FDR}}_{0}(t)\leq\alpha$
for all $t\in\mathbb{N}$, while restricting each threshold $\alpha_{i}$
to be a function of only the first $s_{i}$ test statistics $P_{1},\dots P_{s_{i}}$,
is to set 
\begin{align}
\alpha_{t} & =\left\{ \alpha\left(1\vee\vert\mathcal{R}_{s_{t}}\vert\right)-\sum_{\{i:s_{i}<s_{t}\}}\alpha_{i}\right\} \frac{\pi}{n_{s_{t}}},\label{eq:lord-format}
\end{align}
where $n_{s_{t}}=\sum_{i=1}^{\infty}\mathbf{1}(s_{i}=s_{t})$ is the
number of parameters determined at time $s_{t}$, and $\pi\in[0,1]$
is again a user-specified constant. The summation in Eq (\ref{eq:lord-format})
captures all of the threshold parameters that have been specified
(i.e., planned) before stage $s_{t}$. Thus, at every time $s_{t}$
when thresholds are specified, we ensure that the sum of all thresholds
specified so far ($\sum_{\{i:s_{i}\leq s_{t}\}}\alpha_{i}$) is no
more than $\alpha\left(1\vee\vert\mathcal{R}_{s_{t}}\vert\right)$.
This, in turn, ensures that $\widehat{\text{FDR}}_{0}(t)\leq\alpha$
for all $t\in\mathbb{N}$. 

Likewise, to ensure that $\widehat{\text{FDR}}_{\lambda}(t)\leq\alpha$
while maintaining the restriction that each $\alpha_{i}$ is a function
only of $P_{1},\dots P_{s_{i}}$, we can prompt the user for a prespecified
sequence $\{\lambda_{i}\}_{i=1}^{\infty}$, and then set $\alpha_{t}=\min\{\lambda_{t},\bar{\alpha}_{t}\}$,
where
\begin{align}
\bar{\alpha}_{t} & =\left\{ \alpha\left(1\vee\vert\mathcal{R}_{s_{t}}\vert\right)-\sum_{\{i:s_{i}<s_{t}\}}\frac{\alpha_{i}\mathbf{1}(\lambda_{i}<P_{i}\text{ or }s_{t}<i)}{1-\lambda_{i}}\right\} \left(1-\lambda_{t}\right)\frac{\pi}{n_{s_{t}}}.\label{eq:saff-format}
\end{align}
For each threshold $\alpha_{i}$ specified before time $s_{t}$, the
summation in Eq (\ref{eq:saff-format}) conservatively captures the
contribution of $\alpha_{i}$ to the FDR estimate $\widehat{FDR}_{\lambda}(t)$.
By conservative, we mean that if $P_{i}$ is known by time $s_{t}$
(i.e., $i\leq s_{t})$, then the summation includes the contribution
of $\alpha_{i}$. If $P_{i}$ is not known by time $s_{t}$ (i.e.,
$s_{t}<i$), then the summation includes an upper bound on the contribution
of $\alpha_{i}$. As in Eq (\ref{eq:lord-format}), at any time $s_{t}$
at which thresholds are specified, we see that the set of thresholds
specified so far, $\{\alpha_{i}:s_{i}\leq s_{t}\}$, is guaranteed
to satisfy $\sum_{\{i:s_{i}\leq s_{t}\}}\frac{\alpha_{i}\mathbf{1}(P_{i}>\lambda_{i})}{1-\lambda_{i}}\leq\alpha\left(1\vee\vert\mathcal{R}_{s_{t}}\vert\right)$.
It follows that $\widehat{\text{FDR}}_{\lambda}(t)\leq\alpha$ for
all $t\in\mathbb{N}$.  

Unlike \citet{Zrnic2021-jj}, we do not assume that the specification
times $\{s_{i}\}_{i=1}^{\infty}$ are monotonic in the sense that
$s_{i}\leq s_{i+1}$. That is, we do not necessarily require that
sets of information used to define consecutive parameters form a filtration.
We may, for example, define $\alpha_{1}$ and $\alpha_{3}$ a priori,
but then define $\alpha_{2}$ based on $P_{1}$ (i.e., $(s_{1},s_{2},s_{3})=(0,1,0)$).
This flexibility may be relevant when the timing of how hypotheses
are observed is not fully in the analyst's control.

We show in the supplementary materials that all of the updating rules
described in this section satisfy Condition \ref{cond:my-mono}. These
rules continue to satisfy Condition \ref{cond:my-mono} if we replace
$\pi$ with a sequence of stage-specific, user-defined tuning parameters
$\{\pi_{s_{t}}:t\in\mathbb{N}\}$, which may be useful if either some
hypotheses warrant higher spending than others, or if the hypothesis
sequence is finite and analysts wish spend more aggressively in the
final stages of testing. In the process of demonstrating Condition
\ref{cond:my-mono}, we also develop iterative, computationally efficient
versions of Eqs (\ref{eq:lord-format}) and (\ref{eq:saff-format}).

\section{SIMULATIONS\label{sec:Simulations}}

Here, we illustrate our FDR control result using simulation. Based
on the setup used by \citet{Ramdas2018-qu}, we simulate a vector
of $t_{\text{max}}=500$ normally distributed variables $(Z_{1},\dots,Z_{t_{\text{max}}})\sim N(\mu,\Sigma)$,
where $\mu=(\mu_{1},\dots,\mu_{t_{\text{max}}}$) is a vector of mean
parameters and $\Sigma$ is a covariance matrix defined in detail
below. For each statistic, our null hypothesis $H_{i}$ is that $\mathbb{E}(Z_{i})=0$,
and our p-value $P_{i}=\Phi(-Z_{i})$ is the (unadjusted) result of
a one-sided test of $H_{i}$. In each simulated sample, we select
a random subset of the parameters $(\mu_{1},\dots,\mu_{t_{\text{max}}})$
to be zero, and assign the remaining mean parameters to be 3. Let
$\pi_{1}$ denote the proportion of mean parameters that are equal
to 3, i.e., the proportion of false nulls. 

We define $\Sigma$ according to a block-covariance structure with
block size denoted by $n_{\text{batch}}$, and within-block covariance
$\rho$. As in Section \ref{subsec:Conditional-Superuniformity},
we define $b_{i}$ to be the block label for the $i^{th}$ test statistic.
We define the each element of $\Sigma$ as follows. 
\[
\Sigma_{ij}=\begin{cases}
1 & \text{if }i=j\\
\rho & \text{if \ensuremath{i\neq j} and }b_{i}=b_{j}\\
0 & \text{otherwise}.
\end{cases}
\]

We simulate all combinations of $n_{\text{batch}}\in\{1,5,10,50\}$;
$\rho\in\{0.3,0.6\}$; and
\[
\pi_{1}\in\{0,0.02,0.04,0.06,0.08,0.1,0.2,0.3,0.4,0.5\}.
\]
For each combination, we run 1000 iterations. 

In each simulated sample, we apply the LORD and SAFFRON implementations
described in Section \ref{subsec:planning-ahead-implementation} (see
also \citealp{Zrnic2021-jj}). We require each threshold $\alpha_{i}$
to be chosen based only on the test statistics from previous batches.
That is, we set $s_{i}=\max\{i':b_{i'}<b_{i}\}$, so that $\mathcal{F}_{i-1}=\{P_{i'}:b_{i'}<b_{i}\}$,
and $P_{i}\perp\mathcal{F}_{i-1}$. Here, we can see that CS holds
from the fact that $\mathbb{P}(P_{i}\leq u\vert\mathcal{F}_{i-1})=\mathbb{P}(P_{i}\leq u)=u$
whenever $i\in\mathcal{H}_{0}$. We assign the tuning parameter $\pi$
to be $\min\{1,n_{\text{batch}}\times0.01\}$.

\subsection{Results}

\begin{figure}

\begin{centering}
\includegraphics[width=0.6\columnwidth]{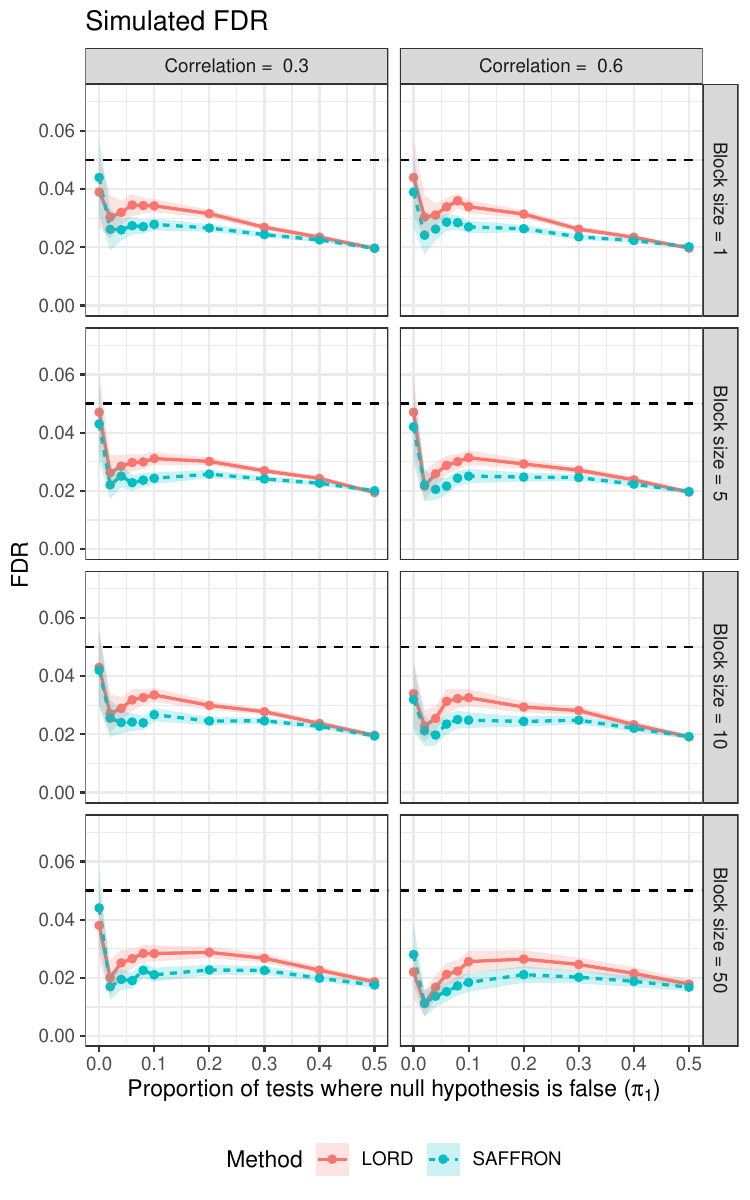}
\par\end{centering}
\caption{\label{fig:Simulated-false-discovery-global}Simulated False Discovery
Rates. We generate z-statistics with a block-correlation structure,
and generate p-value thresholds based only on the p-values from previous
blocks (as in \citealp{Zrnic2021-jj}). Shaded ribbons show $\pm$
2 Monte Carlo standard errors, defined as $\sqrt{1000^{-1}Var(\text{FDP}(t_{\text{max}}))}$.
The dashed line represents our desired alpha level.}
\end{figure}

Figures \ref{fig:Simulated-false-discovery-global} shows the results
of our analysis. We see that, in every scenario tested, LORD and SAFFRON
control $\text{FDR}(t_{\text{max}})$ at the appropriate rate.

\section{CONCLUSION}

Where SAFFRON, LORD and alpha investing were previously only shown
to control FDR under independence, we show that they additionally
control FDR under positive, local dependence.

Although our work focuses on controlling FDR, this should not be taken
as an implicit, blanket endorsement of FDR over mFDR. On the one hand,
\citet{Javanmard2018-rz} argue that FDR carries a more easily understood
interpretation than mFDR. On the other, mFDR may be more easily applicable
within a decision theory framework \citep{Bickel2004-ef}, and may
more naturally allow for decentralized control of the proportion of
false positives across an entire scientific literature (\citealp{Fernando2004-un};
see also \citealp{Van_den_Oord2008-zq,Zrnic2021-jj}). Further research
into online control for both metrics, as well as control for FWER,
remains vital (see also \citealp{Weinstein2020-gv,Tian2021-ha}).

\bibliographystyle{apalike}
\bibliography{_repeated-analysis}

\end{document}